\documentclass[
  aps,
  prc,
  reprint,
  superscriptaddress,
  nofootinbib,
  longbibliography
]{revtex4-2}

\usepackage{prc_qgp_cnn_style}
\usepackage[normalem]{ulem}

\hypersetup{
  colorlinks=true,
  linkcolor=black,
  citecolor=black,
  urlcolor=blue,
  pdftitle={CNN-Based Online Trigger for QGP Event Selection }
}

\begin{document}

\title{CNN-Based Online Trigger for Quark--Gluon Plasma Event Selection}
\author{Olga Soloveva}
 \email{soloveva@itp.uni-frankfurt.de}
\thanks{These authors contributed equally to this work.}
\affiliation{GSI Helmholtzzentrum f\"ur Schwerionenforschung GmbH, Planckstra{\ss}e 1, 64291 Darmstadt, Germany}
\affiliation{Institut f\"ur Theoretische Physik, Johann Wolfgang Goethe University, Max-von-Laue-Str. 1, 60438 Frankfurt, Germany}
\author{Artemiy Belousov}
\thanks{These authors contributed equally to this work.}
\affiliation{Department of Computer Science and Mathematics, Goethe University Frankfurt, Frankfurt am Main, Germany}

\affiliation{Institut f\"ur Theoretische Physik, Johann Wolfgang Goethe	University, Max-von-Laue-Str. 1, 60438 Frankfurt, Germany}

\author{Ivan Kisel}%
\affiliation{Frankfurt Institute for Advanced Studies (FIAS), Ruth Moufang Str. 1, 60438 Frankfurt, Germany}
\affiliation{Department of Computer Science and Mathematics, Goethe University Frankfurt, Frankfurt am Main, Germany}
\affiliation{GSI Helmholtzzentrum f\"ur Schwerionenforschung GmbH, Planckstra{\ss}e 1, 64291 Darmstadt, Germany}
\affiliation{Helmholtz Research Academy Hessen for FAIR (HFHF), GSI Helmholtz	Center for Heavy Ion Physics. Campus Frankfurt, 60438 Frankfurt, Germany.}

\author{Elena Bratkovskaya}%

\affiliation{GSI Helmholtzzentrum f\"ur Schwerionenforschung GmbH, Planckstra{\ss}e 1, 64291 Darmstadt, Germany}
 \affiliation{Institut f\"ur Theoretische Physik, Johann Wolfgang Goethe	University, Max-von-Laue-Str. 1, 60438 Frankfurt, Germany}
\affiliation{Helmholtz Research Academy Hessen for FAIR (HFHF), GSI Helmholtz	Center for Heavy Ion Physics. Campus Frankfurt, 60438 Frankfurt, Germany.}

\date{\today}

\date{\today}

\begin{abstract}
Modern high-rate experimental environments require that rare physics signatures be identified in real time from a continuous stream of reconstructed events, while operating under stringent data throughput and storage constraints. In this context, we present a convolutional-neural-network-based trigger concept for the selection of events associated with quark–gluon plasma (QGP) formation. Events are encoded as compact multidimensional histograms of reconstructed particle content, incorporating particle species, momentum magnitude, and angular information.
A central challenge for such approaches is the possible dependence on the underlying event generator and its physics assumptions. To address this issue, the performance of the CNN is first evaluated within the Parton--Hadron--String Dynamics (PHSD) framework, where direct microscopic QGP-related labels are available. As an independent validation step, the same event representation and network architecture are applied to simulations performed with the Ultra-relativistic Quantum Molecular Dynamics (UrQMD) transport model, which provides a distinct description of the collision dynamics. Finally, cross-checks between PHSD and UrQMD are carried out to assess the stability of the learned response against model-dependent effects and to quantify the model-transfer robustness of the proposed trigger concept.
For deployment in realistic analysis pipelines, a lightweight C++ inference package (ANN4FLES) is employed at the physics-analysis stage after tracking and topology reconstruction. 
The performance is quantified along the full transition from idealized transport-model information to reconstructed data. For Au+Au collisions at 30~AGeV, the classification accuracy decreases from 95.1\% on generator-level PHSD events to 83.7\% after full reconstruction, which remains sufficient for online event selection. In addition, SHAP-based interpretability analysis is used to identify the dominant particle and phase-space contributions to the network decision.
\end{abstract}

\maketitle

\section{Introduction}

The study of strongly interacting matter under extreme conditions remains a central objective of relativistic heavy-ion physics. Heavy-ion collisions provide access to the QCD phase diagram over a broad range of temperatures and baryon densities, including the region where the onset of deconfinement, the nature of the phase transition, and possible critical phenomena are still under active investigation. Since the hot and dense stage of the reaction is not directly observable, information on the possible formation of a quark--gluon plasma (QGP) has to be inferred from experimentally accessible final-state observables after the full nonequilibrium evolution of the collision.

This situation naturally motivates event-by-event classification methods that can identify QGP-sensitive patterns from the reconstructed particle content of heavy-ion reactions. For experimental applications, however, predictive performance alone is not sufficient. A practically useful trigger or analysis classifier must be fast, robust against detector and reconstruction effects, and as independent as possible of a specific event generator. Otherwise, the network may primarily learn model-specific correlations rather than generic signatures of deconfined matter. From this perspective, model independence is not an auxiliary property but a central requirement for experimental applicability.

The present work addresses this problem by developing a convolutional-neural-network (CNN)-based trigger concept that operates on compact multidimensional histograms of reconstructed particles. The key question is whether such a classifier can learn stable QGP-sensitive structures that remain useful under model transfer and detector-level constraints. If this is achieved, the method becomes relevant not only for one particular simulation setup, but as a general strategy for fast online or quasi-online event selection in high-rate heavy-ion experiments
such as the Compressed Baryonic Matter (CBM) experiment at FAIR \cite{Friman:2011zz}, where rare probes of strongly interacting matter at high net-baryon density have to be selected in a free-streaming, software-based analysis chain. In this framework, the First-Level Event Selector (FLES) performs online reconstruction and physics selection, which makes lightweight machine-learning inference particularly attractive. In particular, the ANN4FLES framework was developed as a fast C++ neural-network package for use in the CBM online environment, and recent studies have demonstrated its application both to a CNN-based QGP trigger and to neural-network-based selection tasks inside the reconstruction chain. These developments indicate that the proposed trigger concept is technically compatible with fast online event selection and can contribute to a physics-driven reduction of the output data stream while preserving events of enhanced interest for QGP-related analyses \cite{Sergeev:2020fir,Belousov:2023fqd,Kisel:2025lambda}.

Although CBM provides the most direct application, the physics motivation is broader. A model-independent trigger concept based on reconstructed hadronic final states can be beneficial for other present and future facilities wherever high-rate data taking, rare-event enrichment, or rapid event tagging are required. In the United States, this applies in different ways to the RHIC heavy-ion program and, more generally, to the developing data-intensive environment around the Electron--Ion Collider (EIC). While the EIC is not a heavy-ion QGP facility in the same sense as CBM or RHIC, it represents a major future nuclear-physics experiment whose scientific reach relies on advanced detector concepts, large-scale data processing, and flexible analysis strategies. In that sense, transferable machine-learning methods developed for fast and robust event characterization are of broader methodological relevance beyond a single experiment~\cite{Accardi:2016,EICYellowReport:2022}.

Machine-learning methods have recently been explored as tools for extracting nontrivial correlations from high-dimensional final-state data in relativistic heavy-ion  collisions~\cite{Pang:2020pkl,Zheng:2024dlh}. In particular, CNN-based studies have shown that momentum-space particle distributions can encode information about the QCD equation of state and the nature of the QCD transition~\cite{Pang:2016vdc,Du:2020pmp,Kvasiuk:2020qdd,Mallick:2022alr}. 
More broadly, machine-learning techniques are increasingly used in studies of strongly interacting matter, including lattice field theory, heavy-ion phenomenology, and QCD phase-structure investigations~\cite{Boyda:2022nmh,Zhou:2023pti,He:2023eod}. In a CBM-oriented application, PointNet-based event classification was proposed as an
equation-of-state meter for Au+Au collisions at FAIR energies, demonstrating that machine-learning
methods can retain sensitivity to dense-matter properties under conditions relevant for the CBM
experiment~\cite{OmanaKuttan:2021}. These developments demonstrate that neural networks can identify nonlinear correlations in many-particle final states that are difficult to access through manually designed observables alone. 

The present work extends this direction from offline event characterization toward a trigger-oriented application. We develop a CNN-based method for event-by-event selection of QGP-related events from compact multidimensional representations of reconstructed particle content. It is important to note that the present approach does not employ a physics-informed neural network, nor are explicit physics constraints imposed at the level of the loss function or network architecture. Instead, the CNN is trained directly on multidimensional event representations constructed from the final-state particle content. Therefore, the physics sensitivity emerges from correlations learned by the network from the simulated event samples themselves, rather than from externally prescribed theoretical constraints. 

In the present study, supervised training and validation are performed using events generated with 

two independent transport approaches \cite{Bleicher:2022kcu} -- 
the Parton--Hadron--String Dynamics (PHSD) ~\cite{Cassing:2008sv,Cassing2009,Bratkovskaya2011,Moreau:2019vhw,Jorge:2025wwp} and Ultra-relativistic Quantum Molecular Dynamics (UrQMD) \cite{Bass:1998ca,Bleicher:1999xi,Steinheimer:2011ea,OmanaKuttan:2022the}. 

PHSD provides a fully microscopic off-shell description of relativistic heavy-ion collisions from the initial nonequilibrium stage to the final hadronic freeze-out. It includes hadronic and partonic degrees of freedom, dynamical deconfinement, a strongly interacting QGP phase, hadronization, and the subsequent hadronic expansion. This makes PHSD particularly suitable for assigning event-level labels related to the presence and strength of a partonic phase. In contrast, UrQMD employs an on-shell microscopic transport description for the hadronic degrees of freedom, whereas the QGP phase is modeled macroscopically through hydrodynamic evolution in its hybrid formulation.
This microscopic versus macroscopic realization of the partonic phase increases the difficulty of the CNN classification task. However, it also provides a stringent test of robustness: UrQMD serves as an essential hadronic baseline for determining whether the network learns QGP-sensitive structures rather than generator-specific features. Consequently, cross-model validation is a central element of the present trigger concept, not merely a technical consistency check.

The main objectives of this work are therefore threefold: first, to construct an event representation suitable for CNN inference in heavy-ion collisions; second, to determine whether the classifier learns physically meaningful discriminants associated with QGP formation; and third, to evaluate the stability of the learned decision function under model transfer and under the transition from generator-level to reconstructed events. In this way, the study is aimed not only at a CBM-specific implementation, but at establishing a more general framework for fast and robust machine-learning-assisted event selection in present and future nuclear-physics experiments.

The paper is organized as follows.
In Sec.~II, we describe the methodological framework of the study, including the transport-model framework, the definition of QGP-related labels, the construction of the multidimensional event representation, the CNN architecture, and the SHAP-based interpretability procedure. 
Section~III presents the results. 
We first discuss the physical interpretation of the learned CNN response, then evaluate the training and validation performance on PHSD- and UrQMD-based data sets, followed by a systematic cross-model validation between the two transport approaches. 
The same section also examines the transition from idealized theoretically simulated information to reconstructed detector-level events and discusses the corresponding implications for CBM/FLES-oriented online deployment. 
Finally, Sec.~IV summarizes the main conclusions and outlines future applications of the proposed CNN-based trigger concept for fast online event selection and for model-independent applications in current and future experiments.

\section{Methodology: event representation, \CNN{} architecture, and training setup}
\label{sec:method}

\subsection{Physics event samples and labels}
\label{subsec:samples}

\subsubsection{ PHSD }

The Parton--Hadron--String Dynamics (PHSD) approach provides a microscopic off-shell transport description of relativistic heavy-ion collisions based on the Kadanoff--Baym equations in first-order gradient expansion~\cite{Cassing:2008sv,Cassing2009,Bratkovskaya2011,Moreau:2019vhw,Jorge:2025wwp}. It describes the full nonequilibrium evolution of the reaction, from the initial hard scatterings and string formation through a dynamical deconfinement transition to a strongly interacting quark--gluon plasma (QGP), followed by hadronization and the subsequent hadronic expansion stage. In the hadronic sector, PHSD is essentially equivalent to the Hadron--String Dynamics (HSD) approach, with baryonic and mesonic interactions treated within established hadronic transport theory. 

The partonic phase is described by the Dynamical QuasiParticle Model (DQPM), in which quarks and gluons are represented as off-shell effective quasiparticles with temperature- and chemical-potential-dependent masses and widths constrained by lattice-QCD thermodynamics~\cite{Linnyk2016,Moreau:2019vhw}. In its extended formulation at finite baryon chemical potential, the DQPM also provides medium-dependent partonic cross sections, mean fields, and transport coefficients, enabling a consistent description of strongly interacting matter away from $\mu_B=0$~\cite{Moreau:2019vhw}. 

The transition between hadronic and partonic matter is determined locally by the energy density, while hadronization is implemented dynamically during the expansion through covariant transition rates as the system evolves back into the hadronic regime~\cite{Cassing2009,Bratkovskaya2011}. 
Recent developments have further improved the determination of the microscopic quasiparticle properties by constraining DQPM parameters, such as parton masses and widths, through machine-learning-assisted extraction from lattice-QCD thermodynamics~\cite{Soloveva2024}. 

Due to unified treatment of hadronic and partonic degrees of freedom, PHSD has been successfully applied to p+A and A+A collisions over a broad range of beam energies, from SPS to LHC, including hadronic and electromagnetic observables~\cite{Linnyk2016}. For the present work, PHSD is particularly suitable because it generates fully dynamical event-by-event collision histories with a nontrivial and fluctuating degree of QGP formation. It therefore provides a physically grounded framework for constructing labeled datasets for supervised learning, where the CNN is trained on final-state event information while the target labels are inferred from the underlying microscopic transport evolution.

In PHSD, the onset of the partonic phase is controlled by the local energy density. When the density exceeds a critical value of the order of $\varepsilon_c\sim0.4\,\mathrm{GeV/fm^3}$, the medium is described in terms of partonic degrees of freedom; when the system expands and cools below this threshold, hadronization restores a hadronic description. This microscopic separation makes it possible to tag QGP-origin particles and to define event-level descriptors of the QGP content.

Besides the binary label, PHSD provides quantities such as the QGP energy fraction and the number of particles originating from the QGP phase. An integral QGP-strength parameter $Rat_{qgp}$ or $R_{i}$  can be written as
\begin{equation}
R_{i} = \int_0^{t_f} R(t)\,\frac{E^{\mathrm{QGP}}(t)}{E^{\mathrm{tot}}(t)}\,dt,
\label{eq:Ri}
\end{equation}
where $E^{\mathrm{QGP}}(t)$ and $E^{\mathrm{tot}}(t)$ denote the partonic and total energy, respectively. These labels enable not only binary event classification but also multi-task learning with additional regression targets.

\begin{figure}
    \centering
    \includegraphics[width=1.0\linewidth]{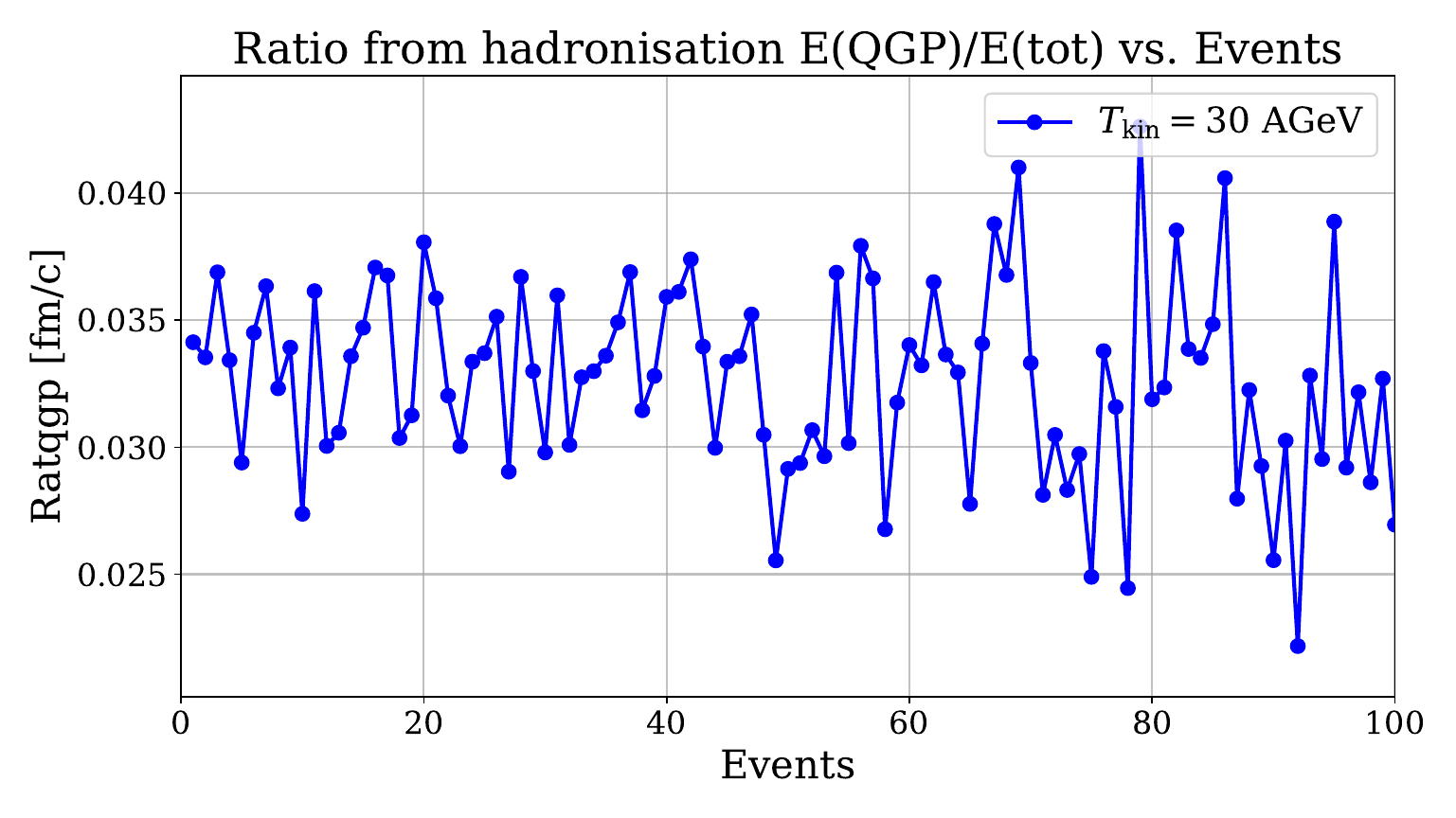}
    \caption{\label{fig:rat27}  The QGP energy fraction from the PHSD as a function of time in central (impact parameter $b = 2.2$ fm) Au+Au at $T_{kin} = 30$ AGeV.}
    \label{fig:qgp_rat}
\end{figure}

The relation between the PHSD evolution and hydrodynamic descriptions has been studied explicitly in Ref.~\cite{Xu:2017pna}, where PHSD was compared to viscous hydrodynamics using the same non-equilibrium initial conditions, an equivalent lattice-QCD-based equation of state, and identical transport coefficients. These studies showed that, despite stronger event-by-event nonequilibrium fluctuations in PHSD, the ensemble-averaged bulk evolution and collective observables are close to the hydrodynamic results~\cite{Xu:2017pna}. This similarity is physically consistent with the fact that the PHSD partonic phase is a strongly interacting QGP described within the DQPM, with an equation of state constrained by lattice QCD and thus close to that used in hydrodynamic approaches~\cite{Cassing:2008sv}.

For the supervised classification task, the PHSD event sample is separated into
two classes. In \textit{Case~1}, the event contains a partonic phase and is therefore
assigned to the QGP class. These events are characterized by nonzero QGP-related
quantities, such as the QGP energy fraction, the integral QGP-strength parameter
$R_i$, and the number of particles originating from the QGP phase,
$N_{\rm qgp}$. In \textit{Case~2}, no partonic phase is formed during the dynamical
evolution of the collision, and the event is assigned to the no-QGP class. In this
case the corresponding QGP-related quantities vanish or remain below the
selection threshold used for defining QGP-positive events.

This separation provides a direct microscopic target for the CNN training. The
network is therefore trained to distinguish events with QGP formation from events
without QGP formation using only the final-state particle information encoded in
the multidimensional input representation. In addition to the binary QGP/no-QGP
classification, the PHSD labels allow the CNN response to be tested against
continuous event-level quantities, such as $R_i$, $N_{\rm qgp}$, and the impact
parameter $b$.


\subsubsection{ URQMD }
To explore the robustness of the CNN approach beyond the PHSD description, we employed the Ultra-relativistic Quantum Molecular Dynamics (UrQMD) model~\cite{Bass:1998ca,Bleicher:1999xi,Steinheimer:2011ea,OmanaKuttan:2022the}
in two different variations. UrQMD provides a microscopic description of heavy-ion collisions in terms of hadronic and string degrees of freedom and is therefore well suited as an independent reference framework for testing whether the classifier response is tied to PHSD-specific correlations or to more general final-state event structures. In the present study, two UrQMD realizations, denoted as {\it Case~1} and {\it Case~2}, are used. They probe complementary aspects of the reaction dynamics and provide independent validation conditions for the CNN.

In {\it Case~1}, we use the core--corona implementation of the UrQMD hybrid approach~\cite{Steinheimer:2011ea}. In this formulation, the system is separated at the transition to the hydrodynamic stage into a dense, approximately equilibrated core and a dilute corona. The core component is evolved hydrodynamically, whereas the corona remains in the microscopic transport description. Since the hydrodynamic core represents the high-density and strongly interacting part of the collision system, it can be regarded phenomenologically as the region where QGP-like conditions may develop. In the dataset used here, particles originating from the hydrodynamic core are explicitly identified at the particle level and can therefore be distinguished from particles emitted from the corona. This information provides a physically motivated validation target for the CNN, analogous in purpose to the PHSD tagging of particles associated with the partonic phase.

In {\it Case~2}, UrQMD is supplemented with a chiral mean-field equation of state~\cite{OmanaKuttan:2022the}. This realization is designed to study the compression stage of heavy-ion collisions and the sensitivity of the reaction dynamics to the underlying equation of state. It provides access to the time evolution of baryon density, temperature, and pressure and is particularly relevant in the beam-energy range $E_{\rm lab}=1$--$10A$~GeV, where compressed baryonic matter and possible phase-transition effects are of central interest~\cite{OmanaKuttan:2022the}. In contrast to {\it Case~1}, where the validation is based on the separation between dense-core and dilute-corona emission, {\it Case~2} probes the stability of the CNN response under a modified dynamical evolution and equation-of-state implementation.

The UrQMD study therefore serves as an independent test and validation of the CNN strategy. Training and validation within the UrQMD samples allow us to verify whether the same event representation and network architecture can reconstruct physically relevant event characteristics in a transport framework different from PHSD. In addition, the comparison of the CNN response between PHSD and UrQMD provides a cross-model test of the learned decision function. In this way, the UrQMD analysis makes it possible to assess whether the classifier extracts robust information from final-state particle distributions, rather than relying only on the microscopic QGP-labeling scheme available in PHSD.

For the purpose of the present machine-learning classification, UrQMD
Case~1 is therefore treated as the dense-medium, or QGP-like, ON class,
whereas UrQMD Case~2 is treated as the corresponding OFF reference class.
This terminology should not be understood as an event-by-event microscopic
QGP label identical to the PHSD one; rather, it defines an operational
classification target within UrQMD, where the presence or absence of a
hydrodynamically evolved core distinguishes the two samples.
Importantly, the two UrQMD samples are not intended to test different
equations of state. They are used to test the sensitivity of the CNN to
the presence of a hydrodynamically evolved dense core versus a purely
transport evolution within the same underlying equation-of-state setup.
 
Taken together, {\it Case~1} and {\it Case~2} provide complementary validation conditions. {\it Case~1} tests the sensitivity of the network to final-state features associated with a dense, hydrodynamically evolving core, while {\it Case~2} tests the stability of the response under changes of the equation of state and compression dynamics. Their combined use is therefore essential for evaluating the model-transfer robustness and physics interpretability of the proposed CNN-based QGP trigger.

It is important to note that, the PHSD results support a core--corona-like interpretation of the fireball evolution, but the underlying physics content differs from that of the UrQMD hybrid implementation. In the UrQMD hybrid model, the system is explicitly separated into an equilibrated, dense core and a dilute, nonequilibrated corona by means of a local density criterion at the transition to hydrodynamics; the core is then evolved hydrodynamically, while the corona remains in the transport description. Within that framework, the core--corona separation improves the description of strange-particle ratios, their beam-energy dependence, and centrality-dependent yields, and it also affects flow observables, indicating that only a fraction of the system can be regarded as approximately equilibrated, especially at lower beam energies \cite{Steinheimer:2011ea}. In PHSD, by contrast, no external core--corona partition is imposed. The coexistence of a hot partonic core and a surrounding hadronic corona emerges dynamically from the microscopic off-shell transport evolution on a cell-by-cell basis. In particular, the local temperature $T$ and baryon chemical potential $\mu_B$ are evaluated in each space-time cell, and the partonic interaction cross sections are determined accordingly \cite{Moreau:2019vhw,Soloveva:2020mdi}. 

The resulting transverse profiles show that the hottest central region corresponds to the QGP phase, whereas the surrounding lower-density matter remains hadronic, thus yielding a natural core--corona-like structure. Accordingly, in PHSD the ``core'' is most naturally identified with the deconfined QGP domain, while the ``corona'' denotes the neighboring hadronic environment; this should be distinguished from the UrQMD hybrid picture, where the core is defined primarily through the applicability of a near-equilibrium hydrodynamic treatment rather than directly through the local microscopic phase content \cite{Steinheimer:2011ea,Moreau:2019vhw}.

\subsection{Input observables and event-image construction}
\label{subsec:inputs}

\subsection{\CNN{} architecture}
\label{subsec:cnn_architecture}
\begin{figure}
    \centering
    \includegraphics[width=0.95\linewidth]{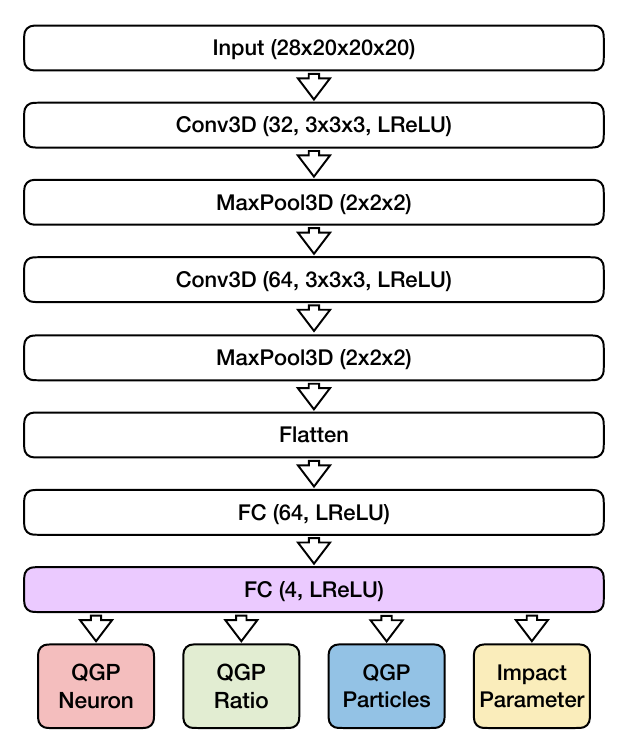}
    \caption{Architecture of the 3D CNN model used in this work. Event tensors of dimension $28\times20\times20\times20$ are processed by two 3D convolutional layers with max-pooling and a fully connected layer. The four output nodes encode the QGP Neuron, QGP ratio, QGP-particle content, and impact parameter.}
    \label{fig:placeholder}
\end{figure}
The CNN architecture employed in this work follows the multi-output design
introduced in the PhD thesis of A.~Belousov~\cite{Belousov:2025nkw}. Each
accepted event is transformed into a four-dimensional histogram representation
with input size $28\times20\times20\times20$. The first dimension corresponds
to the particle-species channel, while the remaining three dimensions encode the
binned kinematic information of the particles: the absolute momentum
$|\mathbf{p}|$, the polar angle $\theta$, and the azimuthal angle $\phi$.

The 28 particle species are selected from the simulated event record by requiring
that a given species appears at least once in 1000 events, thereby reducing
statistical noise from very rare particles. For each selected particle species, the
kinematic variables are discretized into 20 bins. The angular variables are divided
into equal intervals, whereas the absolute momentum is binned logarithmically in
order to provide a denser representation at low momenta, where most particles
are concentrated. Thus, the CNN processes each collision as an event-level
multidimensional image that contains both the particle composition and the
phase-space structure of the final state. This compact representation preserves
the information relevant for QGP-related event classification and is well suited
for fast convolutional processing.
In contrast to the earlier
classification-only implementation with two output neurons for QGP-off and
QGP-on discrimination~\cite{Belousov:2023fqd}, the present network uses a shared
convolutional backbone followed by a four-neuron output head that simultaneously
predicts the QGP-related observables considered in this analysis.

The feature extractor consists of two three-dimensional convolutional blocks.
The first block contains a Conv3D layer with $32$ filters and kernel size
$3 \times 3 \times 3$, followed by a LeakyReLU activation and a MaxPool3D
operation with pooling size $2 \times 2 \times 2$. The second block has the
same structure, but with $64$ convolutional filters of size
$3 \times 3 \times 3$, again followed by LeakyReLU and MaxPool3D with pooling
size $2 \times 2 \times 2$. Thus, the network contains two convolutional
layers, two pooling stages, and LeakyReLU nonlinearities throughout the shared
feature-extraction part~\cite{Belousov:2025nkw}.

After the second pooling layer, the feature maps are flattened and passed to a
fully connected layer with $64$ neurons and LeakyReLU activation. The final
dense layer contains four output neurons, corresponding to the target quantities
used in the present model, namely the QGP classification-related response, the
QGP ratio, the number of QGP particles, and the impact parameter. In the
current implementation, this final layer forms a joint multi-output head rather
than two separate classification logits as in the earlier architecture reported
in Ref.~\cite{Belousov:2023fqd}. The resulting network therefore combines event
classification and regression within a single shared representation
learning framework~\cite{Belousov:2025nkw}.

\subsection{SHAP Values}
Shapley Additive Explanations (SHAP)~\cite{Nohara2019, Shrikumar:2017DeepLIFT} provide an interpretable framework for the analysis of machine-learning models by assigning additive attribution scores to the input features contributing to a given prediction. In this approach, the classifier output is decomposed into feature-wise contributions, which enables a quantitative assessment of how individual observables influence the network response for a specific event, while also allowing one to identify the dominant patterns governing the global behavior of the model.

In the present study, SHAP is applied to a convolutional neural network trained on PHSD events for the classification of QGP and non-QGP events. The SHAP analysis demonstrates that the network response is governed by physically meaningful event characteristics rather than by purely algorithmic correlations.In particular, the largest normalized contributions are associated with strange hadrons and antibaryon-related features, indicating that these observables carry substantial discriminating power for the identification of events containing a deconfined phase.

These results show that SHAP provides a useful tool for relating the internal decision structure of the CNN to the underlying collision dynamics. Such an interpretation is important for assessing the physical robustness of the classifier and for identifying which event features are most relevant for QGP-sensitive triggering in heavy-ion collisions.

Analyzing the SHAP values projected onto particle species leads to the following observations:
\begin{itemize}
  \item The magnitude of the SHAP value reflects the importance of a given particle species for the classification: values close to zero indicate a negligible contribution.
  \item Because the plotted observable is the mean absolute SHAP value, it measures the overall relevance of each particle species for the classifier, irrespective of the sign of its contribution.
  \item The SHAP values are normalized by the particle yield of each species, which makes the contribution of rare particles more clearly visible.
  \item This representation is useful for assessing the impact of rare-particle reconstruction, since losses in the reconstruction efficiency of such species may noticeably reduce the classifier performance.
\end{itemize}

\begin{figure}[htb]
\centering
\includegraphics[scale=0.6]{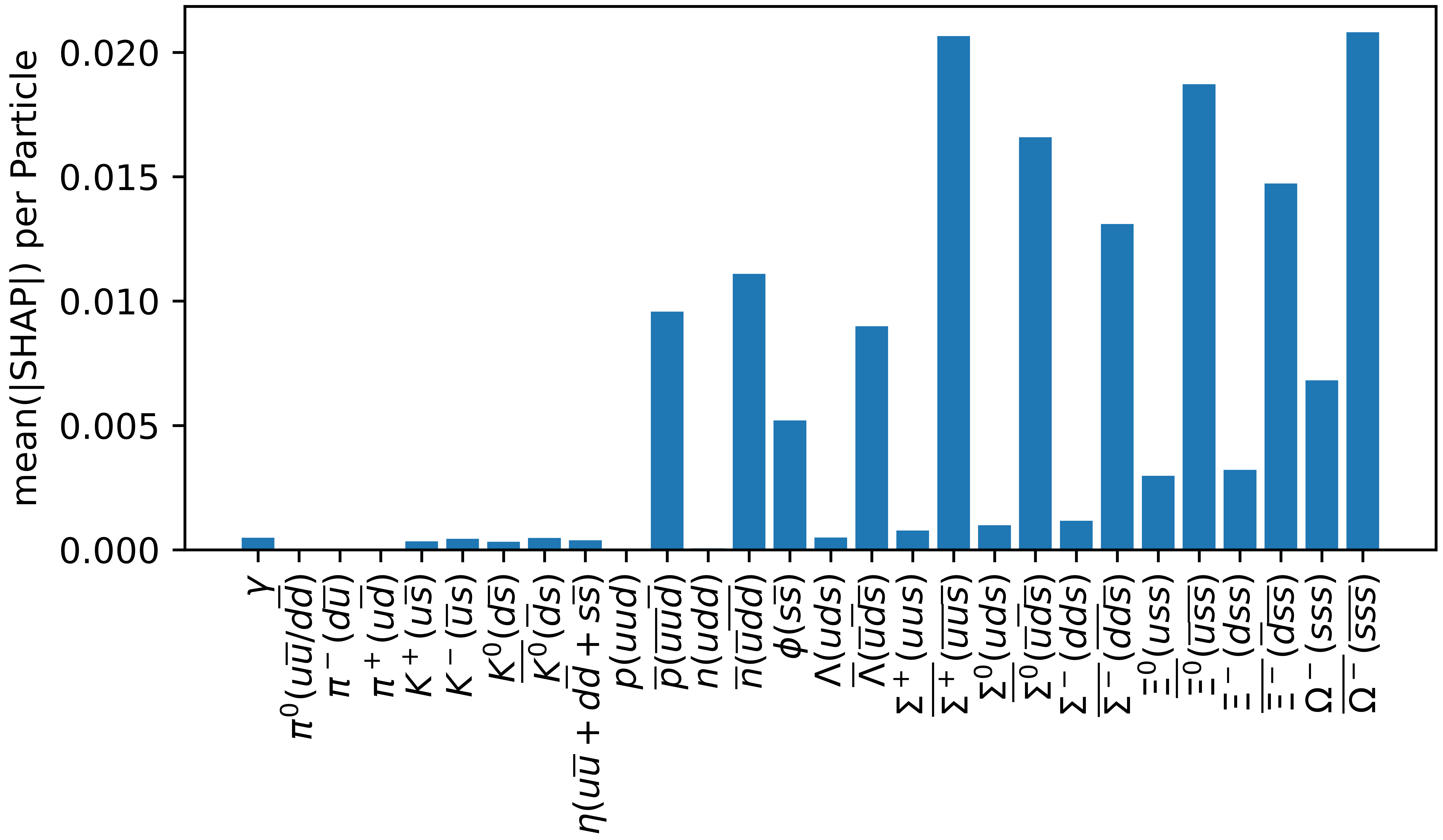} 
\caption{
    Normalized mean absolute SHAP contribution for each particle species, averaged over the analyzed event sample. Larger values indicate a stronger relative influence on the CNN classification.} 
\label{fig:shap_strange}
\end{figure}

Figure~\ref{fig:shap_strange} shows the mean absolute SHAP value per particle species, averaged over the event sample. This quantity measures the average contribution of each particle type to the CNN output, independently of sign, and therefore provides a ranking of their overall importance for the classification task. The distribution exhibits a clear hierarchy of feature relevance, with the largest contributions arising from strange hadrons and, in particular, from antibaryons.

The SHAP analysis indicates that multi-strange and strange antibaryons, such as $\bar{\Sigma}^{+}$, $\bar{\Sigma}^{0}$, $\bar{\Xi}^{0}$, $\bar{\Xi}^{-}$, and $\bar{\Omega}^{-}$, as well as the $\Omega^{-}$, provide the strongest contributions to the QGP-related output of the model. In contrast, light mesons and non-strange particles show substantially smaller average SHAP values. This pattern suggests that the network primarily relies on strange-particle production and antibaryon-related information when identifying events associated with a deconfined phase.

These results are physically meaningful, since enhanced strangeness production has long been discussed as one of the characteristic QGP-sensitive signatures in relativistic heavy-ion collisions, including in the STAR critical assessment of the evidence for QGP formation at RHIC~\cite{STAR:2005gfr}. At the same time, Fig.~\ref{fig:shap_strange} should be interpreted as a feature-importance analysis rather than as a direct measurement of strangeness enhancement in the conventional yield-ratio sense. The dominance of strange particles and antibaryons in the SHAP ranking therefore supports the interpretation that the CNN exploits relevant QGP-sensitive correlations in the final-state particle composition rather than arbitrary correlations in the data. Importantly, this sensitivity is not imposed through an explicit strangeness-enhancement observable or through physics-informed constraints, but emerges from the correlations learned by the CNN directly from the multidimensional final-state particle distributions.

\section{Results}
\label{sec:results}
We use the impact parameter $b$ as a benchmark observable because it is directly available from the transport-model output and can be unambiguously compared with the CNN prediction. While $b$ is not a direct QGP signal, it controls the collision geometry and is correlated with multiplicity, energy deposition, and the probability of forming a dense partonic region. Therefore, an accurate reconstruction of $b$ provides a robust validation that the CNN has learned physically meaningful event features.


\begin{figure}[htbp]
    \centering
    \includegraphics[width=0.98\linewidth]{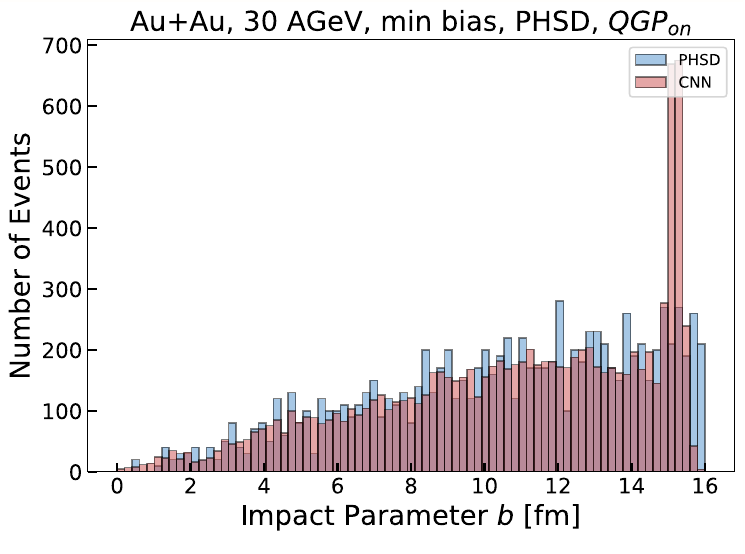}
    \caption{Impact-parameter distribution for PHSD minimum-bias Au+Au events at $30~A$GeV. The reference PHSD labels are compared with the corresponding CNN outputs. }
\label{fig:phsd_b_distribution}
\end{figure}

\begin{figure}[htbp]
    \centering
    \includegraphics[width=0.98\linewidth]{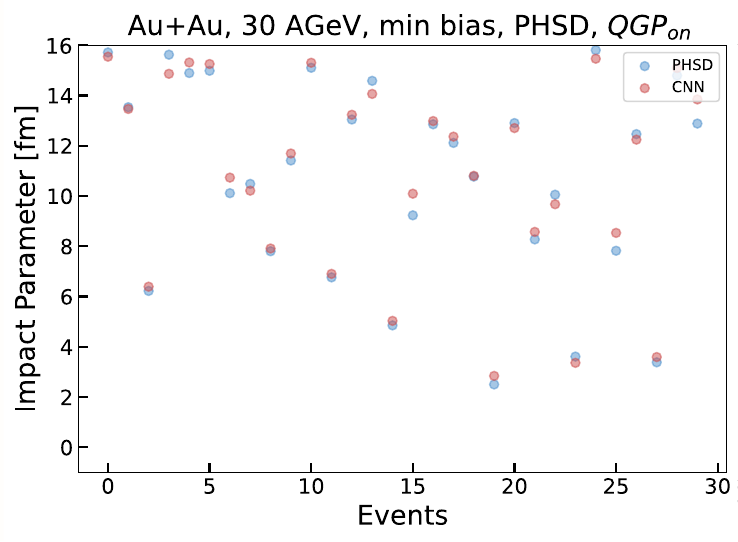}
    \caption{Event-by-event comparison of the impact parameter $b$ for PHSD events in Au+Au collisions at 30~AGeV. The PHSD labels are compared to the CNN predictions for individual events. }
\label{fig:phsd_b_events}
\end{figure}

\begin{figure}[htbp]
    \centering
    \includegraphics[width=1.0\linewidth]{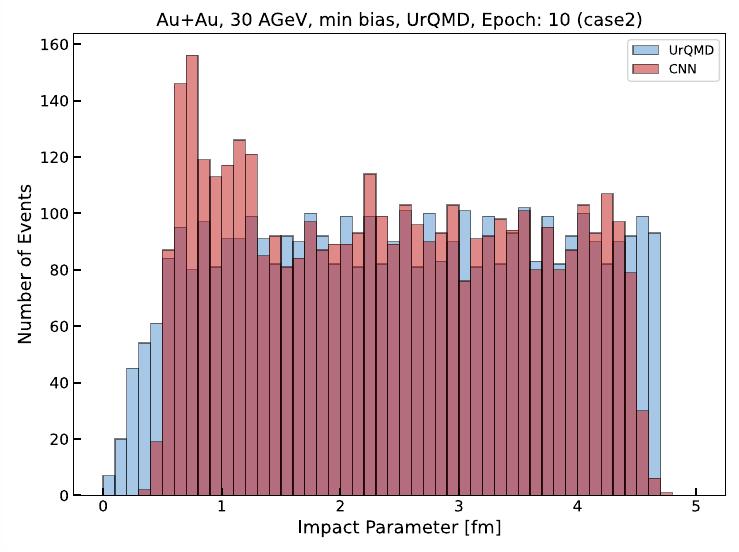}
    \caption{Distribution of the impact parameter $b$ for UrQMD events in Au+Au collisions at 30~AGeV for Epoch~10 (case2). The histogram compares the UrQMD labels with the corresponding CNN outputs. }
\label{fig:urqmd_b_distribution}
\end{figure}

\begin{figure}[htbp]
    \centering
    \includegraphics[width=1.0\linewidth]{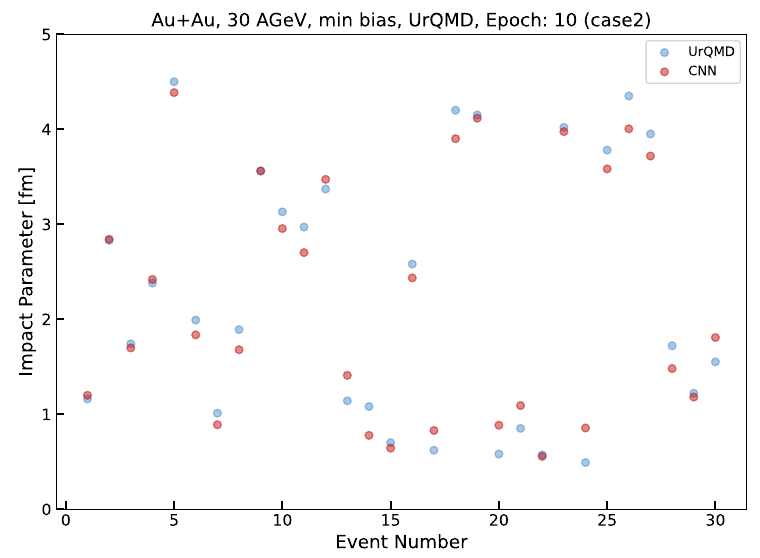}
    \caption{Event-by-event comparison of the impact parameter $b$ for UrQMD events in Au+Au collisions at 30~AGeV for Epoch~10 (case2). The UrQMD labels are compared to the CNN predictions for individual events. }
\label{fig:urqmd_b_events}
\end{figure}

\subsection{Interpretation of learned physics features}
\label{subsec:shap_results}

To relate the CNN response to physically interpretable event characteristics, we
analyzed the trained classifier with Shapley Additive Explanations (SHAP).
Within the DeepExplainer framework, the SHAP tensor was computed for all
network outputs using a balanced background sample of 1000 events, containing
equal numbers of QGP and non-QGP events, and a test sample from a fixed event
class. For the present classifier, this yielded a tensor of dimension
$2 \times 28 \times 20 \times 20 \times 20$, corresponding to the two output
neurons and the full four-dimensional input representation. In the visualization
used here, the SHAP information was compressed by summing over the non-fixed
coordinates, such that the resulting distributions represent contributions
projected onto particle species.
Machine-learning-based impact-parameter and centrality determination has a long history in heavy-ion physics, starting from early neural-network applications~\cite{Bass:1994ee,David:1995,Bass:1996je} and extending to modern deep-learning approaches based on point-cloud and image-like event representations~\cite{OmanaKuttan:2020brg, OmanaKuttan:2021,Li:2020vay,Li:2021}.

The SHAP analysis shows that the classifier decision is not dominated by a
single local feature, but by a distributed pattern of contributions across the
event representation. For events containing a QGP phase, positive SHAP values
for the QGP output neuron identify those regions of the input space that support
the correct classification, whereas near-zero values indicate features with
little influence on the decision. Owing to the binary structure of the network,
the SHAP patterns associated with the QGP and non-QGP output neurons are found
to be approximately mirrored, as expected for a two-class classifier with
anti-correlated outputs.

A central observation is that the normalized SHAP representation enhances the
visibility of contributions from rare particle species. This is important for the
physics interpretation, because species that are subleading in raw abundance can
still carry substantial discriminating power for the QGP decision. In this
sense, the network response is sensitive not only to the overall event activity
but also to the composition of the produced hadronic final state. The SHAP
results therefore indicate that the classifier exploits physically meaningful
differences in species-resolved event structure rather than relying on trivial
global normalization effects alone.

At the same time, the SHAP results documented in the thesis are presented after
projection onto particle type. Consequently, they allow a robust interpretation
in terms of species composition and distributed event structure, but they do not
by themselves isolate the relative importance of multiplicity, baryon stopping,
transverse activity, or geometric topology as separate driving factors. A more
differential decomposition in rapidity, transverse momentum, or spatial
coordinates would be required to make such statements quantitatively. Within the
scope of the present analysis, the conservative conclusion is that the CNN
learns correlated many-particle patterns associated with QGP-sensitive event
composition and uses these patterns consistently in the binary discrimination
task. 

\subsection{Training and validation on PHSD and UrQMD events}
\begin{figure*}[htbp]
    \centering
    \includegraphics[width=0.88\linewidth]{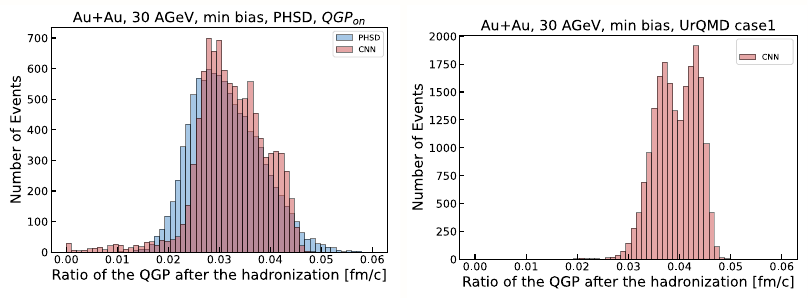}
    \caption{
    Restricted transfer comparison of the CNN-predicted integrated QGP-strength response for PHSD and UrQMD events. The left panel shows the PHSD QGP-on sample, for which microscopic QGP labels are available, while the right panel shows the UrQMD Case1 sample associated with the dense hydrodynamic/core region. The UrQMD sample is used as a proxy for an unlabeled experimental data set, illustrating whether the CNN response remains meaningful when no direct event-by-event QGP label is available.}
\label{fig:cross_model_b}
\end{figure*}

The PHSD-based training provides the reference case for the present CNN trigger study, because the target quantities are defined directly from the microscopic transport history of the event. In this setup, the network is trained and validated on minimum-bias Au+Au collisions at $30~A$GeV, including events with and without a partonic phase. In addition to the binary QGP classification label, each event contains continuous PHSD labels, such as the QGP ratio $R_i$, the number of particles originating from the QGP phase, $N_{\rm QGP}$, and the impact parameter $b$. The CNN therefore performs a combined classification and regression task: it identifies QGP-related events and simultaneously reconstructs physically relevant event-level quantities.

Figure~\ref{fig:phsd_b_distribution} shows the distribution of the impact parameter $b$ obtained from the PHSD reference labels and from the corresponding CNN output. The close agreement between the two histograms demonstrates that the network reproduces the global collision-geometry distribution with good accuracy. This result is nontrivial, since $b$ is not used as an input observable, but has to be inferred indirectly from the final-state particle composition and kinematic correlations. The comparison therefore shows that the CNN extracts event-geometry information encoded in the multidimensional particle representation, rather than simply learning the average multiplicity scale.

Figure~\ref{fig:phsd_b_events} provides an event-by-event comparison for individual PHSD events. The CNN predictions follow the PHSD labels over a broad range of impact parameters, confirming that the agreement observed in the inclusive distribution is also preserved at the single-event level. Such an event-wise comparison is important for a trigger application, because online selection must operate on individual collisions rather than on ensemble-averaged observables. The PHSD validation therefore demonstrates that the updated CNN architecture is able to reconstruct the main event characteristics, including the impact parameter and QGP-related quantities, with stable accuracy under minimum-bias conditions. This establishes PHSD as a controlled baseline for training a QGP-sensitive trigger with direct access to microscopic partonic information.

The UrQMD-based study provides an independent validation of the same CNN architecture in a different transport-model environment. In contrast to PHSD, UrQMD does not provide the identical microscopic definition of a dynamically formed QGP phase. Instead, the UrQMD datasets used in this analysis contain event-level information associated with dense/core-like regions and with the event geometry. The CNN architecture is therefore adapted to reconstruct the corresponding UrQMD target quantities, in particular the number of QGP-related or dense-region particles and the impact parameter $b$. This test is essential for determining whether the network response remains meaningful when the event generator and the underlying microscopic dynamics are changed.

Figure~\ref{fig:urqmd_b_distribution} shows the impact-parameter distribution for the UrQMD sample, comparing the UrQMD labels with the CNN predictions. The similarity of the two distributions indicates that the network preserves the overall event-geometry structure also for UrQMD events. This agreement shows that the learned representation is not restricted to the PHSD label construction, but can also be applied to an independent hadronic/hybrid transport description after conversion to the same input format.

Figure~\ref{fig:urqmd_b_events} presents the corresponding event-by-event comparison for UrQMD events. The CNN predictions remain correlated with the UrQMD labels for individual events, demonstrating that the network reconstructs the event characteristics not only statistically, but also on an event-wise basis. The stability of the response for both UrQMD cases confirms that the CNN captures robust features of the final-state particle distribution, including information related to collision geometry and dense-region particle production. We therefore conclude that the trained network response is not generator-specific, but remains physically meaningful when applied to an independent transport-model dataset.

A decisive test for a QGP-oriented CNN trigger is whether the learned response remains meaningful when the network is transferred between different transport models. This question is essential because experimental data do not provide direct event-by-event QGP labels, and a classifier trained on a single generator may otherwise learn model-specific correlations rather than robust physical event structures. Following the strategy discussed in Chapter~7 of Ref.~\cite{Belousov:2025nkw}, the cross-validation is therefore performed using both PHSD and UrQMD event samples.

In the PHSD sample, the QGP-related target quantities are defined directly from the microscopic transport history, where the dynamical formation of a partonic phase is explicitly available. In contrast, the UrQMD samples provide an independent transport-model environment with different microscopic assumptions and without an identical PHSD-like definition of the QGP phase. For this reason, the comparison between PHSD and UrQMD is not expected to yield a one-to-one correspondence of all output quantities. Instead, the relevant question is whether the CNN preserves a stable and physically interpretable response when applied across models.

Figure~8 shows a restricted transfer test of the CNN output associated
with the integrated QGP-strength target. This figure should not be
interpreted as the complete cross-validation between all event classes.
A full cross-model validation requires the simultaneous comparison of
PHSD QGP-on, PHSD QGP-off, UrQMD Case~1, and UrQMD Case~2 events, and is
discussed separately in Sec.~III~C. The more limited purpose of
Fig.~8 is to compare the CNN response for event samples that contain a
QGP-like or dense-medium component: PHSD QGP-on events and UrQMD
Case~1 events.

The horizontal axis represents the CNN-predicted value of the
time-integrated QGP-strength descriptor $R_i$ defined in Eq.~(1), not an
instantaneous QGP fraction at the final time. In PHSD, the microscopic
QGP energy fraction $E_{\rm QGP}(t)/E_{\rm tot}(t)$ is accumulated over
the partonic stage of the reaction. Since the integration is performed
over time, $R_i$ has units of fm/$c$. After hadronization the
instantaneous QGP fraction is zero; however, the integrated
quantity remains finite because it records the earlier presence of a
partonic phase during the dynamical evolution. Thus a value
$R_i\simeq 0.03$~fm/$c$ should be understood as a small integrated
QGP-strength proxy in the adopted normalization, rather than as a
post-hadronization QGP fraction.

For UrQMD Case~1 there is no identical microscopic PHSD-like definition
of the QGP energy fraction. The UrQMD sample is therefore used as an
independent dense-medium proxy, based on its hydrodynamic/core component.
The fact that the CNN output lies in a comparable range indicates that
the network response is not tied exclusively to PHSD-specific labels, but
retains sensitivity to correlated final-state structures associated with
dense-medium formation. This result should be regarded as evidence for a
transferable QGP-like response, not as a direct proof of event-by-event
QGP identification in UrQMD.

In this comparison, the UrQMD events are used as a proxy for an unlabeled
experimental sample. While labels are available at the simulation level, they are
not used to define a PHSD-like event-by-event QGP truth during the inference
step. This reflects the situation in real experiments, where no direct microscopic
QGP label is accessible for individual events. The fact that the CNN response for
UrQMD follows a pattern similar to the PHSD result in Fig.~\ref{fig:cross_model_b}
therefore supports the interpretation that the network has learned transferable
final-state correlations rather than merely reproducing generator-specific
training labels.
The observed separation of the PHSD and UrQMD classes indicates that the CNN response is not restricted to a single transport-model realization. In particular, the network trained on PHSD-like QGP information also produces a meaningful ordering of the UrQMD classes, with the QGP-like response associated predominantly with Case1 and the no-QGP-like response with Case2. This behavior supports the interpretation that the CNN learns correlated final-state patterns connected with dense-medium formation, particle composition, and event geometry. The result should therefore be understood as evidence for model-transfer robustness, rather than as a strict model-independent proof of QGP identification.

This cross-validation provides an important consistency check for the proposed trigger concept. Since PHSD and UrQMD differ in their treatment of the collision dynamics, perfect agreement between the two models is neither expected nor required. The relevant outcome is that the CNN retains discriminating power under model transfer. This demonstrates that the classifier captures nontrivial event features that are common to different descriptions of heavy-ion dynamics and supports its further development as a model-robust online selection tool for QGP-sensitive events.

\subsection{Cross-model validation: PHSD versus UrQMD}
\label{subsec:cross_model_validation}

A central question for any QGP-oriented event classifier is whether the learned
decision function reflects robust physical event structure or merely model-specific
correlations. To address this point, we performed a systematic cross-model
validation using events generated with the PHSD and UrQMD transport
approaches. The corresponding tests were organized in four steps:
(i) training and testing on PHSD,
(ii) training and testing on UrQMD,
(iii) training on PHSD and testing on UrQMD, and
(iv) training on UrQMD and testing on PHSD.
This sequence allows one to separate the intrinsic performance of the network on
a fixed event generator from its transferability across models with different
microscopic dynamics.

The PHSD-based training provides the reference case in which the target labels
are defined directly through the partonic content of the event. In this setup, the
classifier is trained to distinguish QGP-on from QGP-off events and, in the
multi-output formulation, to reconstruct additional event-level quantities such as
the number of QGP particles, the integral QGP-strength parameter, and the
impact parameter. Since PHSD contains an explicit partonic phase with
dynamical deconfinement and hadronization, this test defines the baseline
performance of the network when the labeling is directly tied to the microscopic
transport evolution.

A second reference test is obtained by training and testing on UrQMD data. In
the present setup, two UrQMD event classes are considered. Case1 contains
information on particles associated with the dense hydrodynamic core, whereas
case2 provides the corresponding non-core reference. After conversion of particle
identifiers to the common input representation, the same CNN can be trained on
UrQMD events in a fully analogous way. This benchmark establishes the
classifier performance within a purely hadronic/hybrid event generator framework
and provides a necessary comparison to the PHSD baseline.

The essential model-independence test is then obtained by transferring the
network between the two transport descriptions. In the first transfer direction,
the classifier is trained on PHSD events and subsequently evaluated on UrQMD
events that were not used during training. In this case, the PHSD QGP-on and
QGP-off classes define the learned decision boundary, while the UrQMD case1
and case2 samples are used only for inference. Within the dissertation study,
this transfer test showed that the classifier continues to separate the two UrQMD
classes, with the QGP-like response associated predominantly with case1 and the
non-QGP-like response with case2. This observation indicates that the network
does not rely exclusively on model-specific features of PHSD, but captures a set
of event characteristics that remain relevant when applied to an independent
transport description.

The inverse transfer, namely training on UrQMD and testing on PHSD, provides
the complementary validation. Conceptually, this test is equally important,
because it probes whether a decision function learned from the hydrodynamic-core
structure in UrQMD can identify events with explicit partonic content in PHSD.
If the classifier response remains stable in this direction as well, the conclusion
is stronger: the learned representation is controlled primarily by common
final-state patterns associated with a dense, collective stage of the reaction,
rather than by artifacts tied to a particular implementation of the underlying
dynamics. In practice, the two transfer directions should therefore be viewed as a
paired test of transport-model robustness.

The physical interpretation of these cross-model studies is straightforward.
PHSD and UrQMD differ substantially at the microscopic level: PHSD contains
explicit partonic degrees of freedom and dynamical hadronization, whereas the
UrQMD samples employed here represent either a core--corona or a chiral
equation-of-state realization without an identical event-by-event QGP label.
Accordingly, perfect transfer is neither expected nor required. What is relevant
is whether the classifier preserves a statistically significant separation power when
moved from one model to the other. A stable response under such transfer implies
that the network has learned nontrivial correlations in particle composition and
kinematics that are shared across different descriptions of high-density
heavy-ion dynamics.

These cross-model tests therefore provide more than a technical validation of the
network. They quantify the degree to which the classifier output can be
interpreted as a model-robust proxy for QGP-sensitive event structure. This is
particularly important for any future application to experimental data, where the
true microscopic label is not available and the utility of the method depends
directly on its insensitivity to the choice of training generator.


In summary, the four-step validation procedure
\begin{align}
\mathrm{PHSD}\!\to\!\mathrm{PHSD},\quad
\mathrm{UrQMD}\!\to\!\mathrm{UrQMD}, \nonumber\\
\mathrm{PHSD}\!\to\!\mathrm{UrQMD},\quad
\mathrm{UrQMD}\!\to\!\mathrm{PHSD}.
\end{align}
provides a direct measure of model dependence in the classifier response.
The first two cases determine the within-model reference performance, whereas
the last two isolate the genuinely transferable part of the learned representation.
In this sense, cross-model validation is a necessary condition for interpreting the
CNN output as a physically meaningful indicator of robust event structure rather
than as a generator-specific discriminator.
\subsection{\CBM{}-oriented performance and deployment aspects}
\label{subsec:cbm}
\begin{figure}[htb]
    \centering
\includegraphics[width=0.5\textwidth]{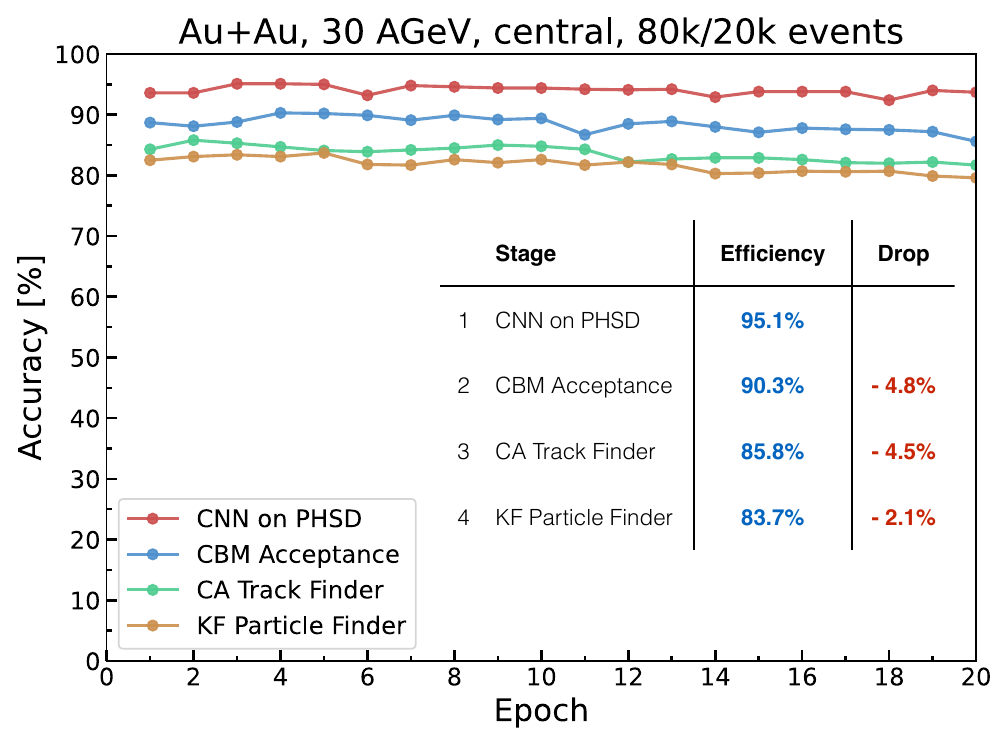}
    \caption{Classification accuracy at different stages of data processing for Au+Au collisions at 30~AGeV. 
    The graph on the left shows the accuracy as a function of training epochs for various datasets. 
    The table on the right summarizes the accuracy at each stage and the drop compared to the previous step.}
    \label{fig:qgp_classification_ch6}
\end{figure}

\label{subsec:cbm}

The practical relevance of the proposed classifier is determined not only by its
performance on idealized Monte Carlo events, but also by its stability under the
detector and reconstruction constraints expected in the \CBM{} online analysis
chain. For this reason, we evaluate the transition from generator-level PHSD
events to reconstructed event information obtained within the \FLES{} framework.
This step is essential for assessing whether the classifier can be used as a realistic
component of a software-based QGP trigger rather than merely as an offline
proof-of-principle study.

Within \CBM{}, the \FLES{} package performs online event reconstruction from
detector signals using a sequence of modules for track finding, track fitting, event
building, particle reconstruction, and physics analysis. In the implementation
considered here, the neural-network inference is naturally placed at the physics
analysis stage, where the reconstructed event content is available in a compact and
experiment-oriented form. In particular, the classifier can be interfaced through
\ANNFLES{}, which receives the reconstructed particle information after the
application of the \CA{} Track Finder, the \KF{} Track Fitter, and, where needed,
the \KFParticle{} Finder for short-lived hadrons. In this way, the trigger decision
is based not on ideal event records, but on the same class of observables that
would be accessible in an actual online environment.

The corresponding input representation is therefore detector-driven. Relative to
the generator-level case, the network no longer receives complete event
information, but a reduced and distorted image of the collision after detector
acceptance cuts, tracking inefficiencies, combinatorial ambiguities, and losses in
topology reconstruction. This feature is important for two reasons. First, it makes
the classifier output directly relevant for online deployment. Second, it provides a
quantitative measure of how much QGP-sensitive information survives the
transition from idealized event records to reconstructed final states.

To evaluate the robustness of the classifier under realistic experimental
conditions, the transport-model output was processed through the CBM detector
simulation and reconstruction chain. In addition to pure Monte Carlo event
information, several increasingly realistic input levels were considered: the CBM
geometrical acceptance, track reconstruction with the CA Track Finder, and full
topology reconstruction with the KF Particle Finder~\cite{Belousov:2025nkw}.
The detector response was simulated within the CbmRoot environment
~\cite{AlTurany:2006CbmRoot} by transporting particles through the CBM setup
with GEANT4~\cite{Agostinelli:2002hh,Allison:2006ve}, followed by
reconstruction within the FLES chain~\cite{Belousov:2025nkw}.

This procedure allows one to quantify the degradation of the CNN performance
when moving from idealized generator-level information to reconstructed event
information affected by detector acceptance, tracking.

The degradation of the classification accuracy through the successive processing
stages is summarized in Fig.~\ref{fig:qgp_classification_ch6}. At the generator
level, the CNN applied directly to PHSD events reaches an accuracy of
\(95.1\%\). After imposing the \CBM{} acceptance, the accuracy decreases to
\(90.3\%\), corresponding to a loss of \(4.8\%\). This reduction reflects the fact that
a fraction of the produced particles does not enter the active detector region owing
to finite geometrical coverage and dead areas.
The CBM acceptance is implemented by applying the polar-angle cut
$2.5^{\circ}<\theta<25^{\circ}$, corresponding to the angular coverage of the
CBM detector system. This range covers the forward rapidity region including
midrapidity at FAIR beam energies~\cite{Senger:2020cje}.
The next step, based on the
\CA{} Track Finder together with Monte Carlo mother-particle information,
reduces the accuracy further to \(85.8\%\), i.e., by an additional \(4.5\%\). This loss is
associated primarily with track-density effects in central events and with
combinatorial ambiguities in the reconstruction of charged-particle trajectories.
Finally, after the inclusion of the \KFParticle{} Finder for the reconstruction of
short-lived particles, the accuracy becomes \(83.7\%\), corresponding to a further
drop of \(2.1\%\). At this stage, the remaining losses are driven mainly by
imperfections in decay reconstruction and by the enhanced combinatorial
background associated with topology finding. The total reduction from ideal PHSD
input to the fully reconstructed chain is therefore about \(11.4\%\).

The rapid increase of the training accuracy in Fig.~\ref{fig:qgp_classification_ch6}
is expected from the mini-batch training procedure and from the relative
homogeneity of the considered event sample. The full dataset contains
$10^4$ event files, which are repeatedly passed through the network during
training. Within a single epoch, the optimizer does not update the CNN weights
only once after processing the complete dataset. Instead, the input sample is
divided into mini-batches, and the weights are updated after each mini-batch.
Thus, already during the first epoch, the network performs several optimization
steps and can rapidly adapt its convolutional filters to the dominant
event-level correlations encoded in the multidimensional input representation.

This fast convergence is further supported by the fact that the events considered
here correspond to central heavy-ion collisions and therefore exhibit a certain
similarity in their global properties, in particular in the overall particle
multiplicity and phase-space occupancy. As a result, the CNN can identify the
most relevant event-level structures already after a limited number of parameter
updates. This explains the fast initial rise of the accuracy observed in the
left panel of Fig.~\ref{fig:qgp_classification_ch6}. The subsequent behavior of
the curves reflects the stability of the learned representation when the input
information is progressively restricted by the CBM acceptance, tracking, and
topology reconstruction effects.

Despite this cumulative degradation, the final performance remains sufficiently
high for online event selection. The result of \(83.7\%\) demonstrates that the
classifier preserves substantial separation power even after realistic detector and
reconstruction effects are taken into account. This is the relevant figure of merit
for a trigger-oriented application, because it quantifies the efficiency with which
QGP-sensitive events can still be identified once the event content is reduced to
experimentally available observables. In this sense, the study indicates that the
CNN does not depend critically on idealized information absent in realistic data,
but instead exploits correlations that survive the full reconstruction chain.

From the deployment point of view, the main advantage of the \ANNFLES{}
implementation is that the inference stage can be incorporated directly into the
existing \FLES{} workflow. Since the classifier acts on a compact histogram-based
representation of reconstructed particles, the computational task is lightweight
compared to full event reconstruction. The natural role of the network is
therefore not to replace reconstruction, but to provide a fast final-stage decision
layer for event enrichment and data-rate reduction. In a trigger scenario, the
selection threshold can be adjusted depending on whether the priority is maximal
QGP efficiency or stronger suppression of background-like events. The resulting
trade-off between trigger efficiency and background rejection can thus be tuned
according to the physics stream and the available bandwidth.

A further important conclusion is the robustness of the approach against
incomplete information. The observed accuracy losses originate from precisely the
effects expected in a realistic detector environment: finite acceptance, dead zones,
tracking uncertainties, particle-identification limitations, multiple scattering, and
the loss of low-momentum particles. The fact that the classifier retains useful
performance after all of these effects are included indicates that the learned event
representation is not tied to a fragile subset of perfectly reconstructed particles.
Rather, it is sufficiently redundant to remain informative under moderate detector
smearing and partial topology loss. This feature is essential for any future
application to real data, where the event record is necessarily incomplete and
distorted relative to the microscopic truth.

Overall, the results shown in Fig.~\ref{fig:qgp_classification_ch6} support the
feasibility of using the CNN classifier as a \CBM{}-oriented online selection tool.
The gradual but controlled reduction in performance from ideal events to the full
reconstruction chain shows that the trigger concept remains operational under
realistic experimental conditions. This makes the proposed classifier a viable
candidate for integration into \ANNFLES{} as part of the \FLES{} physics analysis
stage, where it can contribute to fast event tagging and to the enrichment of data
samples with enhanced sensitivity to QGP-related dynamics.

\section{Conclusions and outlook}
\label{sec:conclusions}

In this work, we have developed and systematically validated a CNN-based trigger
concept for the selection of QGP-related events in heavy-ion collisions. The
approach combines a compact detector-oriented event representation with a
lightweight 3D convolutional architecture suitable for fast inference in realistic
analysis environments. Trained and tested on event samples generated with the
 PHSD and UrQMD transport approaches, and further tested for PHSD events after propagation through the CBM/FLES detector and reconstruction chain, the method provides a
general strategy for identifying QGP-sensitive event structure from reconstructed
final-state information. A central and distinctive feature of the present study is
that the classifier is not formulated as a model-tuned discriminator for a single
transport approach, but is explicitly designed and validated as a model-robust
classification framework. This makes the method relevant not only for one
specific experiment, but more broadly for heavy-ion measurements in which rare
events associated with deconfined or strongly collective matter must be isolated
under realistic detector and computing constraints.

The first principal result is that the CNN identifies QGP-relevant event
patterns with high and practically useful performance. At the generator level,
the classifier exhibits strong discrimination power on PHSD events. More
importantly, this performance remains substantial after the successive inclusion
of realistic detector and reconstruction effects. When the full chain from ideal
PHSD events to reconstructed \CBM{} events is taken into account, including
acceptance effects, track reconstruction, and the reconstruction of short-lived
particles, the classification accuracy decreases from \(95.1\%\) to \(83.7\%\). This
reduction is expected and reflects the loss of information induced by finite
acceptance, imperfect tracking, combinatorial background, and topology-finding
limitations. At the same time, the remaining performance demonstrates that the
classifier exploits persistent and physically meaningful correlations in the
reconstructed final state rather than fragile features tied to idealized event
records. In this sense, the method satisfies a necessary condition for trigger and
event-enrichment applications in heavy-ion experiments: the learned event
representation remains informative after realistic experimental degradation.

The second principal result is the explicit demonstration of transport-model
robustness. Cross-model validation between PHSD and UrQMD shows that the
classifier retains separation power when transferred from one microscopic
description to another. This is a central outcome of the present work. In
heavy-ion collisions, the relation between the transient deconfined stage and the
experimentally accessible hadronic final state is necessarily indirect and depends
on the theoretical description used to generate the training data. Any
machine-learning-based trigger constructed from simulation therefore faces the
fundamental question of model dependence. The results obtained here show that
the classifier response is not exhausted by generator-specific correlations, but is
sensitive to a class of event structures that survives across different microscopic
realizations of the reaction dynamics. This makes the method substantially more
valuable than a model-specific discriminator. It demonstrates that event
selection based on machine learning can be formulated in a way that is not only
efficient, but also physically transferable. From this perspective, the
PHSD--UrQMD comparison is not merely a technical benchmark, but one of the
key scientific results of the study.

A further important aspect of the present work is the formulation of a transferable multi-model CNN strategy, in which different transport-model datasets can be
used within a common network architecture while preserving the physical meaning
of their model-specific targets. This provides a principled way to combine
heterogeneous theoretical input without imposing an artificial one-to-one mapping
between all observables. Such a construction opens the possibility of extending
the classifier to broader and more diverse training ensembles, thereby increasing
its robustness with respect to theoretical uncertainty. In this sense, the present
study establishes not only a concrete trigger implementation, but also a general
framework for model-robust event selection from transport simulations in
heavy-ion physics.

The third principal result is that the classifier admits a physically meaningful
interpretation. The SHAP-based analysis indicates that the network decision is controlled by structured variations in particle composition, with important contributions from strange hadrons and antibaryon-related channels, consistent with expected QGP-sensitive final-state patterns. This is essential for experimental use. A classifier of this
type cannot be regarded only as a numerical black box; its response must be
understood as a proxy for physically relevant event structure. The present
results show that this requirement can be met. The combination of predictive
performance, cross-model stability, and interpretability is therefore one of the
most significant achievements of this work.

Taken together, these results demonstrate that the proposed method has broad
potential for fast event selection and event enrichment in heavy-ion experiments.
Its practical realization within \CBM{} and the \FLES{} chain provides an
important proof of deployment readiness in a high-rate environment. At the same
time, the underlying strategy is not restricted to \CBM{}. Because it is based on
compact reconstructed event information, fast CNN inference, explicit
cross-model validation, and physically interpretable output, it defines a more
general approach for experiments in which rare QGP-sensitive events or events
with pronounced collective dynamics must be identified under limited bandwidth
and substantial background. The method is therefore relevant not only as an
online trigger candidate, but also more generally as a framework for real-time or
quasi-real-time event characterization, sample enrichment, and physics-driven
data reduction in heavy-ion programs.

The broader scientific importance of this result should be emphasized. To our knowledge, this is among the first studies in which a QGP-oriented trigger concept is formulated, quantitatively tested across different transport models, and evaluated through a realistic detector-reconstruction chain while maintaining practical performance. This combination of ingredients is unique. It establishes a
direct connection between microscopic heavy-ion theory, modern deep-learning
methods, and experimentally relevant event selection. In doing so, it shows that
transport-informed machine learning can move beyond offline phenomenological
classification and become a realistic component of the analysis toolkit for future
studies of strongly interacting matter.

The outlook is correspondingly broad. A natural next step is the extension of
the present framework to larger sets of theoretical descriptions, including
additional transport and hybrid approaches, in order to sharpen further the
model independence of the learned representation. Such extensions will help to
identify which parts of the classifier response are genuinely universal and which
remain sensitive to specific assumptions of the underlying dynamics. A second
major direction is the calibration and validation of the method on experimental
data once suitable samples become available. This will allow a direct assessment
of simulation-to-data domain shifts and provide the basis for data-driven
refinement of the classifier response. A third direction is the further optimization
of the network and inference framework for online deployment, including
hardware-aware implementations on heterogeneous CPU/GPU/FPGA
architectures and systematic studies of the operating point for different trigger
efficiency versus background rejection requirements.

In summary, the present work establishes that a CNN-based trigger can identify
QGP-relevant event patterns with high practical performance, that the learned
decision function is physically interpretable, and that its response remains
substantially stable under transfer between different transport models and under
realistic detector-reconstruction conditions. These properties make the proposed
method not only a viable candidate for \CBM{}, but also a conceptually new and
broadly relevant framework for model-robust event selection in heavy-ion
physics. The combination of uniqueness, transport-model robustness, and
deployment readiness demonstrated here opens a realistic path toward the use of
machine-learning-based QGP-sensitive triggers and event classifiers in future
heavy-ion experiments.

\begin{acknowledgments}
The authors thank Marcus Bleicher and Jan Steinheimer for providing the UrQMD data sets used in this study, for valuable discussions, and for their careful reading of the manuscript. 

We are grateful to Iu.~Vasiliev, Akhil Mithran for fruitful discussions.
The authors thank the CBM Collaboration for discussions and for providing the simulation and reconstruction frameworks used in this study. We also acknowledge the support by the Deutsche Forschungsgemeinschaft (DFG) through the grant CRC-TR 211 "Strong-interaction matter under extreme conditions" (Project number 315477589 - TRR 211).
The computational resources for this project were provided by the Center for Scientific Computing of the GU Frankfurt and the Goethe–HLR and GSI green cube.
\end{acknowledgments}

\bibliographystyle{apsrev4-2}
\bibliography{refs}

\end{document}